# Emission dipole orientation reveals dynamic single-molecule interactions with 2D crystals at solvent interfaces

Wei Guo[1,*], Tzu-Heng Chen[1,*], Nathan Ronceray[1], Eveline Mayner[1], Kenji Watanabe[2], Takashi Taniguchi[3], Aleksandra Radenovic[1,*]

[1] *Laboratory of Nanoscale Biology, Institute of Bioengineering (IBI), School of Engineering (STI), École Polytechnique Fédérale de Lausanne (EPFL), Lausanne, Switzerland.*

[2] *Research Center for Electronic and Optical Materials, National Institute for Materials Science, 1-1 Namiki, Tsukuba 305-0044, Japan.*

[3] *Research Center for Materials Nanoarchitectonics, National Institute for Materials Science, 1-1 Namiki, Tsukuba 305-0044, Japan.*

**** Corresponding Emails: wei.guo@epfl.ch, tzu-heng.chen@epfl.ch, aleksandra.radenovic@epfl.ch***

## Abstract

Direct observation of individual fluorescent emitters is essential for studying quantum materials, chemical reactions, and biological systems. However, current single-molecule tracking methods only focuses on the localizations of molecules, overlooking molecular configuration and orientation. In this work, we introduce a high-throughput polarized single-molecule localization microscopy that simultaneously resolves the locations and emission dipole orientations of single fluorescent emitters with nanometer precision. Using the interface between pristine hexagonal boron nitride (h-BN) and an organic solvent as a challenging platform, we capture over $10^5$ fluorescent events and reveal distinct molecular interaction dynamics at room temperature. The measured dipole orientations align with the three-fold ($C_3$) rotational symmetry of the h-BN lattice, and molecular dynamics in the liquid environment can be modulated electrochemically, suggesting a route for on-demand control of quantum emitters. We also find that lateral diffusion at the solid–liquid interface is far more dynamic than that of solid-state emitters. This simultaneous tracking of molecular conformation and photophysics advances the understanding of single-molecule interactions and enables real-time sensing through two-dimensional materials.

# Introduction

Large-scale characterization of fluorescent emitters and the observation of single-molecule conformational dynamics are essential for quantum materials[1,2], chemical reactions[3,4] and biological systems[5,6]. Solid-state quantum emitters have been widely used in computing and sensing[7-9]; and these color centers are usually characterized and observed using optical methods[1,9-11]. However, for other fluorescent emitters that are located at solid-liquid interfaces and are extremely dynamic[12-15], then traditional optical techniques are troubled by bulk measurements and optical diffraction limits (200–300 nm resolution). For example, recent studies have increasingly focused on the hexagonal boron nitride (h-BN) with single molecules[14,16-20], because of its high mechanical strength, atomically flat surface and the ability to slow down the diffusion movement. The wide band gap of h-BN also makes it compatible with fluorescence imaging on the surface compared to graphene and $MoS_2$ substrates[21,22]. Surface defects on h-BN play a critical role not only in solid-state quantum emitters but also in molecular sensing in nanofluidics[13,14], where the quantum emission could also be induced in the solvent environment.

To understand the photophysics of these emitters in complex environments with high throughput, interactions between molecules and pristine h-BN can be studied by single-molecule localization microscopy (SMLM). However, conventional SMLM only provides information about the intensity and localization of the emitters, comprehensive profiling demands additional properties such as spectrum[23-25] and polarization[2,14,26]. Among these properties, the three-dimensional (3D) orientation of the emission dipole is of particular interest, since quantum emitters, which formed by irradiation[27,28], doping[29-31], and etching[32,33], usually result in dipolar emission that are modulated by the external environment. This interest also stems from the compatibility with SMLM without sacrificing temporal and spatial resolution. Although a variety of techniques have been developed for 3D-orientation SMLM (e.g., pupil splitting[34-36], back-focal-plane modulation[37-39], excitation modulation[40,41]) to study the heterogeneity of biological systems, their potential in materials science remains to be fully explored. Dipole orientation provides a window into intrinsic molecular conformations and their interactions with materials[42-44]. Therefore, high-throughput molecular orientation and localization detection enables the transition from “case studies” to quantitative statistical analysis of molecular dynamics.

In this work, we utilize a customized four-pupil polarized single-molecule localization microscopy (pSMLM) to map the nanoscale localization and track the 3D-orientation of fluorescent emitters. This approach correlates molecular orientation and conformation with the crystal lattice in a solvent environment. By optimizing azimuthal accuracy, our four-pupil design specifically adapts to the challenges of the 2D material interface, enabling simultaneous 6D imaging. Our findings reveal that molecular interactions at the liquid-solid interface are regulated by both the h-BN crystal structure and electrochemical conditions, providing insights into the dynamic behavior of emitters in complex surface environments.

# Results

## Experimental concept

To observe non-invasively the molecular dynamics of the h-BN surface in response to the nanoscale environment[14,45], thick h-BN flakes (tens to hundreds of layers) are exfoliated from the high-quality crystals to the coverslip and immersed in organic solvent. We chose acetonitrile as the experimental solvent due to its linear geometry, without tautomer, and highest purity (>99.9%) among commercial organic solvents. The whole sample is then placed in a custom-built chamber and fixed on the inverted pSMLM microscope (Fig. 1a). The 561 nm laser is chosen as an excitation beam and fluorescent emitters are generated at the liquid-solid interface of h-BN flakes. The averaged spectrum of these emitters is 600-700 nm[14,45], but it is still challenging to interpret the behavior of each emitter based on this averaged characteristic and spectrum. These optically excited and sparsely dispersed blinking emitters are presented as diffraction-limited point spread functions (PSF), making them suitable for analysis with single-molecule localization microscopy in the range of nanometers.

Moving beyond the standard 2D Gaussian of PSFs, polarization measurement extracts 3D dipole orientations by analyzing the fluorescence of transition dipole moments. This capability is realized through a custom four-pupil-splitting pSMLM setup (Fig. 1b), designed to facilitate high-throughput characterization of fluorescent emitters at solid-liquid interface. We selected this strategy because it balances acquisition time (number of photons) with localization accuracy more effectively than alternative designs[46-48]. Due to its compact PSF shape, this strategy achieves superior signal-to-noise ratios and the highest azimuthal accuracy compared to other PSF-engineering methods. To meet the needs of molecular dynamics studies at the millisecond scale, in this work, we also extend it to orientation measurements of the azimuth ($\varphi$), polar ($\theta$), and wobble angles ($\alpha$) of emission dipoles (the insert of Fig. 1b) based on the rotational dipole model[49,50] (Supplementary Note 1). The azimuth and polar angles represent the average orientations of the dipole in the plane and out of the plane, respectively, while the wobble angle is the half-cone angle when the molecule is assumed to wobble uniformly inside a hard cone. The fluorescence from the dipoles (red path) is divided by three beam splitters into four channels with different linear polarization states: 0°, 90°, 45°, and 135° separately. Subsequently, four polarized PSFs from the same emission dipole are recorded simultaneously by a single camera and the simulated annealing algorithm is employed to estimate the 3D-orientation by iterating a loss function involving the PSFs and degree of linear polarization (DoLP) map (Supplementary Note 2).

The Fig. 1c shows the super-resolved reaction sites of the liquid-activated emitters on the h-BN surface in high throughput and the nanoscale resolution of SMLM compared to widefield in the zoomed-in diagram. The imaging strategy is similar to point accumulation in nanoscale topography (PAINT), where the transient landing and leaving of fluorescent emitters on h-BN generate a series of single-molecule fluorescence signals. After individual fluorescent emitters are localized by fitting their PSFs, the brightness of a reconstructed spot in the SMLM image represents the frequency in which the emitter was detected at that location (Fig. 1c). Furthermore, these discrete but bright spots are spread throughout the entire flake, so the size of these clusters is related to the photon numbers and dynamics of generated emitters, and it determines the resolution of the SMLM image ($\approx$30 nm in this work, as estimated by the decorrelation algorithm[51] in Supplementary Note 3). Further results on dipole orientation and rotational mobility are shown in Fig. 1d and each parameter can be rendered as a pseudo-color map, showing the ability of the method to reveal wobble, polar, and azimuthal angles. Moreover, Fig. 1e shows two typical experimental polarized-PSFs and their fitting results, for which an empirical interpretation can be stated as follows considering a stationary dipole: The distribution of the light intensity on the four polarization channels will depend on the azimuthal angle, while the presence of the polar angle will lead to shifts and deformations of polarized PSFs. In the case of wobbling dipoles, both features will be degraded. In the following sections we will present a step-by-step description of the molecule-crystal interaction through dipole orientation.

## The breaking of axial symmetry

Directly measuring the structure and dynamics of single-molecule interactions is highly challenging. We propose using dipole orientation as a fast and sensitive method to track changes between different molecular structural states, based on their distinct dipole configurations. Here, the h-BN in the acetonitrile solvent creates fluorescent emission in the visible range through a reversible interaction. At a constant power of $2 \times 10^3$ W cm$^2$ with circularly polarized laser (ellipticity > 97%) as the excitation light, the ability to extract polar and wobble angles allows us to identify three distinct populations instead of two (Fig. 2a) on a flat crystal where no edges were included. The bin width of the 2D heat map (Fig. 2a) is chosen based on numerical simulations of polar-PSFs at different SNRs (Supplementary Note 2.2), which takes care of overestimating orientational accuracy compared to the Cramer-Rao lower bound. Fig. 2b illustrates the orientation of three states in the spherical coordinate and rotational diffusion in a cone. The first state has the smallest polar angle and allows the dipole to wobble in a large conical region due to thermal fluctuations, etc. The second state is almost parallel and firmly locked to the h-BN surface. The third state is slightly different in that it has a larger wobble angle compared to the first, but the polar angle is nearly 90°. Since the $C_3$ symmetry and 2D confinement effect of h-BN, this state suggests that it originates from the binding of fluorescent emitters with rotational mobility, instead of the unrestricted wobbling, at symmetric contact sites on the lattice. Moreover, the angular fluorescence intensity of a fixed emission dipole at the far-field depends on $<\sin(\theta)^2>$. The emitter in state-3 has the best overlap with the excited circularly polarized light in the *X-Y* plane, so the excitation and photon collection efficiency are the highest. However, the emission in state-2 and state-1 gradually weakens as the polar angle decreases. Therefore, the light intensity of three clusters is in ascending order with the polar angle and agrees with the normalized theoretical value, as reflected in the normalized histograms in Fig. 2c.

Both solid-state quantum emitters and fluorescent dyes are regulated by external conditions[2,45,52-54], and our situation is no exception. To further understand the dynamics of the emitters, we employed an electrochemical system[45] while also measuring dynamically the polarization states. Here we use a two-electrode system with a working electrode (indium tin oxide, ITO) and a counter/reference electrode (platinum, Pt). All potentials are herein reported against the Pt reference. The whole sample was scanned with cyclic voltammetry from 0-2 V for 15 cycles to ensure the stability of electrodes and solvent in this working range (Supplementary Note 3). Fig. 2d shows the change in the number of localizations during the three consecutive cycles. Under the application of an increasingly positive electrochemical potential (orange curve), the localization (blue curve) decreases and vice versa, which can be observed by all SMLM techniques but cannot be correlated to the molecular structure[45,52]. But when analyzing each set of emitter states separately, we find that the proportion (abbreviated as P, violet curve) of wobble angles $\alpha > 15°$ (i.e., state-1 and -3, P1+P3) is consistent with the trend of the overall number of localizations. In contrast, the proportion of state-2 ($\alpha < 15°$) increases with increasing voltage, suggesting that emitters with large wobble angles are more easily quenched and vice versa.

Under the condition of no voltage applied (Fig. 2a), the value of P1+P3 is dominates the variation of the overall numbers. However, under the influence of the potential gradient, the share of this fraction is significantly suppressed to less than 50% (Fig. 2d), indicating that the rotational diffusion correlates with more reactivity and is more sensitive to the environmental conditions than that of the state-2. In terms of temporal dynamics of these emitters, although macroscopic changes in the number of localizations respond to the changing electrochemical environment with a phase delay of a few seconds, the gradient (green curve, $dN_emitters/dt$) shows that the emitter does respond immediately when the voltage is reversed. Besides, after stacking the histograms of polar angle for state-2 into a 2D heat map (Fig. 2e), the pixel colors in each row indicate the number density at the current potential. Although the structure of state-2 is more stable, it can be noticed that these dipole orientations still shift slightly in the polar angle. The emitter density during the voltage rise phase produces a significant displacement at 0.7 V, with a peak 40°±1° (dashed line); when the voltage falls back to 0.4 V, the emitter density returns to its original distribution, and the total number of emitters is still lagging behind the potential. On the right we show five typical polar-angle distribution curves and fit them with a double Gaussian model, which clearly shows the shift of the two peaks under electrochemical modulation.

## In-plane orientations of emitters

In this section, we turn our attention to the azimuth angle of each emitter to study how fluorescent emitters interact with the crystal, taking advantage of the well-defined crystal structure of h-BN. All localized emitters are rendered as pseudo-color maps and stacked over time based on the azimuth angle (Fig. 3a). The multiple orientations observed at specific locations on the crystal demonstrate the heterogeneity of emitters, suggesting that multiple closely spaced binding sites cannot be fully resolved because of the resolution limit of SMLM (≈30 nm). The azimuth angles of emitters in the four regions of interests (ROIs) can be plotted as short sticks. Three of the ROIs are sharp crystal edges and one is a line channel. The orientation of these dipoles is not strictly aligned with the crystal edges, where ROIs 1, 3, and 4 are at angles to the boundary, and ROI 2 is oriented coinciding with the direction of the crystal edge. However, when we plot the histogram of the $\Delta\varphi=|\varphi_{edge}-\varphi_{emitters}|$, these angles are well-separated in three groups and are spaced radially apart from each other, indicating a structured arrangement rather than a random distribution. When we compare the box plots in Fig. 3d, the interquartile range of ROIs 4 is significantly smaller than that of ROIs 3 even though the directions of the two regions are almost parallel, which can be explained by reasoning that the line channel confines the emitters more than the open edge. The larger width of the box plot for ROIs 2 is understood in that two peaks exist in Fig. 3c. The respective proportions of the three states in each ROI are P1>P2>P3, while the fraction of the state-3 is almost negligible in these areas (Fig. 3e), compared to the flat area (dash line). This suggests that although crystal edges provide more contact opportunities for molecules (i.e., the edges are brighter than the center in both SMLM and widefield results), rotational binding sites cannot be guaranteed because of the missing symmetry of the h-BN crystal at the edges.

We further examine the cluster of state-2 in the flat crystal area for which the azimuthal angle is most accurate, because the rotation-limited model considers the average orientation of the dipole, and thus dipole wobble will weaken the azimuthal accuracy significantly. The angular distribution in the flat center of the flake exhibits six distinct peaks and can be grouped into two clusters (Fig. 3f), where they are marked in red and orange separately. And when the sample was rotated about 21° (Supplementary Note 5), the position of the azimuthal distribution moved the same as well. Then according to the $C_3$ symmetry of the h-BN structure, the azimuthal orientation of dipoles initially measured in the 0°-180° range can be reconstructed to span the entire 0°-360° range (Fig. 3g) and overlapped with the crystal lattice in Fig. 3h. Remarkably, we now readily find that the azimuthal angle does not strictly follow the h-BN structure, where the averaged orientation is 47° ± 5° (median ± standard deviation) with respect to the lattice and the angle of the two symmetric clusters is 26° ± 7°. Here we show for the symmetry of the emitter orientation with respect to the material lattice with the help of h-BN crystals.

## Molecular dynamics at solid-liquid interface

Interfacial fluorescent emitters are more dynamic than solid-state emitters and 2D h-BN is an ideal platform for single-molecule tracking due to their specific adsorption capacity for molecules and appropriate kinetic temporal scales[14,18,55]. We selected around $2 \times 10^5$ localizations within a 12 × 12 $\mu m^2$ flat region at the center of the crystal, and more than $3 \times 10^4$ trajectories to obtain reliable statistical results. Fig. 4a shows a trajectory map for events longer than four consecutive frames, which are connected based on the localization of random active emitters with nanometric precision (see Methods section on trajectory linking and analysis). These connected events show two main types: random walking on the crystal surface and trapping into bright spots by a fixed location, which is discriminated according to the probability density function (PDF) of the one-dimensional displacements shown in Fig. 4b. These PDFs can always be represented by a superposition of two Gaussian models with increasing delay times ($\tau$). The narrowed Gaussian with a constant standard deviation of 30 nm represents the trapped event, while the broader part with an increased standard deviation represents the trajectories of random walk. The diffusion coefficient of these emitters in trajectory is $7.02\times10^{-14}$ $m^2$ $s^{-1}$. The dwell times of the trapped event and the trajectory follow the double exponential decay as Fig. 4c, and the time constants that correspond to chemisorption are 35 and 68 ms, respectively. The inset shows that the traveling distance of trajectory events is significantly longer than trapped events. However, these emitters do not exhibit significant axial motion because of the attraction of h-BN crystals in Fig. 4d (Supplementary Note 6).

The emitter dynamics so far have been limited to the analysis of localizations. But here we can extend the dimension of molecular diffusion by including orientational information, allowing us to further stratify the trapped events into anchored, jumping, and switching behaviors. Their respective percentages approximately follow the exponential distribution as shown in Fig. 4e and the Fig. 4f shows typical examples of these events. The anchored denotation means that the orientation and mobility of the dipole remain constant when the emitter is continuously activated at a fixed location. The jumping event indicates that the orientation of a dipole changes significantly from one stable state to another during successive activations. Finally, the event of switching means that the emitter, although trapped in a fixed position within the resolution of SMLM, switches back and forth between two equilibrium states. The Fig. 4e shows that the dwell times of the three subgroups are close to each other and their dwell times are on the same order of magnitude. In addition, because the orientation of these emitters is changed from one state to another, then we interpret this as the alternate capture of individual molecules by neighboring localizations rather than the reorientation in identical positions. When the position of each state is marked (stars in the trajectory illustration in Fig. 4f) in the event, it enables the grouping of emitters with similar states, thereby enhancing the SMLM resolution for samples without prior knowledge. This would not be possible without manually altering the environmental conditions, suggesting that this behavior may come from at least two sources. The double localization[56] results of these two demonstration events show that the distances between the two neighboring point sources are 8 ± 3 nm and 11 ± 4 nm respectively. The 2D heat map of the trapped events (Fig. 4g) shows that trapped emitters are mostly belong to state-2, which means the freely moving emitters are more active as states-1 and -3.

Finally, we turn our attention back to the trajectory events, which also include three basic types, as shown in Fig. 4h: (1) the molecule is captured after random walking; (2) captured before leaving and (3) wandering freely on the surface. The azimuth angle is represented by an arrow in the figure, the color of the arrow codes the polar angle, and the arrow length is the wobble angle. The three basic behaviors described above can be further combined into complex behaviors (4) and (5). The trajectory(4) indicates that the same molecule is captured by three different locations consecutively, while trajectory(5) is that a molecule is captured by the same location multiple times. This can be validated by the squared displacement (SD) of Fig. 4i, where the trapped positions are presented as platforms on the curve. In addition, the angular difference between the orientation of the dipole and the direction of movement is uniformly distributed in Fig. 4j, indicating that the movement of the dipole is mainly governed by Brownian motion instead of the crystal lattice and that the transition dipole moment does not affect the macroscopic motion of the molecule. When we review the trajectory shown in Fig. 4h, we find that the trapped emitter seems to have a relatively smaller wobble angle (short arrow) or a larger polar angle (green color of arrow). This further validates the observation in Fig. 4g that the degree of angular freedom is inversely correlated to capture events.

# Discussion

This work applies single-molecule orientation-localization microscopy to investigate the polarization properties of fluorescent emitters at the solid-liquid interface, where molecular dynamics are more complex than in solid-state systems. To address this, we present an end-to-end pipeline, ranging from calibration to data analysis, for tracking orientational and localization dynamics. High-throughput characterization reveals that the dipole orientation reflects molecular conformation and exhibits $C_3$ symmetry relative to the crystal lattice. Our experiments show that the rotational mobility of emitters is highly sensitive to electrochemical conditions, and their 3D-orientation has a significant preference for the h-BN structure. Additionally, the solid-liquid interface offers extensive possibilities for the tracking of single-molecule dynamics: emitters can remain anchored, jump, and move laterally on the h-BN surface. This combination of molecular conformation and dynamics allows us to achieve real-time and large-scale readouts of excited states at the molecular level, which are unmatched by bulk measurements. While this orientation analysis is most effective for linearly polarized emitters (e.g., carbon/oxygen doped defects) rather than high-symmetry centers (e.g., $D_{3h}$ boron vacancies), it remains a powerful tool for detecting symmetry breaking under electric fields or external stress. Ultimately, we aim to advance the understanding of molecular interaction, enhance trace sensing in nanofluidics, and open up possibilities for advanced optical characterization by enabling the simultaneous tracking of molecular conformations and photophysical activity.

# Methods

## Sample preparation

The precision coverslips (CG15XH, Thorlabs) were sonicated in 2% Hellmanex III and rinsed with deionized water for residual impurities before being exposed to oxygen plasma (90 s, 100 W). Pristine h-BN flakes were exfoliated from high-quality crystals by using low-adhesion Nitto tape and then transferred to clean coverslips without pressing them. Samples were left to stand for 24 h to improve transfer efficiency. The sample was mounted on the custom microscope and its surface was completely covered with acetonitrile solvent (99.9% purity). The sample chamber was rinsed twice with the same acetonitrile solvent to clean residues before imaging. The chamber is machined from the polyetheretherketone (PEEK) and it is resistant to organic solvents. We used coverslips coated with an ITO layer (70-100 Ω per square, Diamond Coatings Ltd) and kept electrodes fully immersed in the organic solvent when during electrochemistry experiments.

## Imaging setup

The 561nm laser is modulated into circularly polarized light (97% ellipticity) by a combination of a polarizer (LPVISE100-A, Thorlabs) and a quarter-wave plate (WPQ05M-561, Thorlabs), and then illuminates the sample with normal incidence. The experimental sample is fixed on the nano-positioning systems (Mad City Labs) by a customized sample stage. Fluorescence is collected by a 60×, 1.42 NA oil-immersion objective lens (PLAPON60XO, Olympus), which performs well for polarization imaging. Excitation light around 561 nm wavelength is reflected by the dichroic mirror (Di03-R561-t1, Semrock), and the fluorescence from the sample is further filtered out by a combination of low-pass (750 nm, to remove the fluorescence from boron vacancy) and notch filters (561nm, to clean the reflected light from excitation beam). The variable waveplate (LCC1223-A, Thorlabs) is used for compensating the phase distortion caused by the birefringence and transmission coefficient difference of the first dichroic mirror coating. After the tube lens, a 50:50 non-polarizing beam splitter (BS016, Thorlabs) splits the fluorescence into two channels, and the polarization state of one path is rotated 45° by the half-wave plate (AHWP10M-580, Thorlabs), and then two polarizing beam splitters (PBS251, Thorlabs) split them into four channels. To introduce *z*-axis information and to take full advantage of the polarization properties, the PSF image of one redundant channel is defocus about 150 nm compared to the other three channels. Finally, all beams are reflected by mirrors to sCMOS (Prime-95B, Photometrics) for imaging and the maximum FOV is about 40×40 μm$^2$. In summary, the pSMLM employed in this study achieves precise calibration of polarization distortion through two steps: variable waveplate compensation in the emission path and numerical compensation in post-processing. As a comparison, Rimoli[35] used a four-channel setup with the same focal plane to perform 2D-localization and relied solely on light intensity to calculate the in-plane angle and the in-plane projection of the wobble angle. The effective NA of the objective lens is limited to 1.2 to maintain the accuracy of pure light intensity calculation. However, in our work, axial information coupling and three-dimensional localization are realized by defocusing one redundant channel among the four polarization channels ($I_0+I_{90}=I_{45}+I_{135}$). Finally, the azimuth angle, polar angle, and wobble angle of the dipole were obtained through PSF fitting, and the localization bias caused by the dipole angles was corrected at the same time. Therefore, we can achieve six-dimensional (*x, y, z, φ, θ, α*) simultaneously imaging.

### Imaging detection and estimation algorithms

The imaging strategy was performed in an out-of-focus mode at about 100 nm in the axial axis, and the images of the four channels were registered by affine transformation and localized. Subsequently, after the PSFs of the same dipole in each channel are taken out from the raw data and assigned together, dipoles with empty channels of more than two are deleted, avoiding mislocalization by utilizing intrinsic physical constraints. The azimuthal angle of the dipole is calculated directly from the difference in light intensity of each channel, and the polar and wobble angles are fitted simultaneously by the simulated annealing algorithm.

### Trajectory linking and analysis

When the localization of the emitter in two consecutive frames is less than a given distance threshold, these two emitters will be connected on the temporal sequence. For trapped events, this threshold is 32 nm, and the threshold for randomly walking trajectories is 125 nm. The distance threshold is set according to the full width at half maximum (FWHM≈2.355$\sigma$ for a Gaussian distribution) of the probability distribution of one-dimensional displacements, so the confidence that the event originated from the same emitter is larger than 95.44% (2$\sigma$). The pipeline used to identify the three subcategories of capture events is as follows: The first step is to identify if a reorientation has occurred using outliers and thresholds to determine if it belongs to an anchored event. Subsequently, jumping and switching events are further determined by comparing the difference between the dipole orientations at the beginning and the end of the event sequence.

## Data availability

The data that support the findings of this study are available from the corresponding authors upon request. Source data are provided with this paper.

## Code availability

The code used in the study are available from the corresponding authors upon request.

## References

1 Sutula, M. *et al.* Large-scale optical characterization of solid-state quantum emitters. *Nature Materials* **22**, 1338-1344 (2023). https://doi.org/10.1038/s41563-023-01644-8

2 Kumar, A. *et al.* Polarization Dynamics of Solid-State Quantum Emitters. *ACS Nano* **18**, 5270-5281 (2024). https://doi.org/10.1021/acsnano.3c08940

3 Brinks, D. *et al.* Ultrafast dynamics of single molecules. *Chemical Society Reviews* **43**, 2476-2491 (2014). https://doi.org/10.1039/C3CS60269A

4 Serag, M. F., Abadi, M. & Habuchi, S. Single-molecule diffusion and conformational dynamics by spatial integration of temporal fluctuations. *Nature Communications* **5**, 5123 (2014). https://doi.org/10.1038/ncomms6123

5 Zhao, J. *et al.* Ligand efficacy modulates conformational dynamics of the μ-opioid receptor. *Nature* **629**, 474-480 (2024). https://doi.org/10.1038/s41586-024-07295-2

6 Zhou, W. *et al.* Resolving the Nanoscale Structure of β-Sheet Peptide Self-Assemblies Using Single-Molecule Orientation–Localization Microscopy. *ACS Nano* **18**, 8798-8810 (2024). https://doi.org/10.1021/acsnano.3c11771

7 Aharonovich, I. & Toth, M. Quantum emitters in two dimensions. *Science* **358**, 170-171 (2017). https://doi.org/doi:10.1126/science.aao6951

8 Aharonovich, I., Englund, D. & Toth, M. Solid-state single-photon emitters. *Nature Photonics* **10**, 631-641 (2016). https://doi.org/10.1038/nphoton.2016.186

9 Aharonovich, I., Greentree, A. D. & Prawer, S. Diamond photonics. *Nature Photonics* **5**, 397-405 (2011). https://doi.org/10.1038/nphoton.2011.54

10 Glushkov, E. *et al.* Waveguide-Based Platform for Large-FOV Imaging of Optically Active Defects in 2D Materials. *ACS Photonics* **6**, 3100-3107 (2019). https://doi.org/10.1021/acsphotonics.9b01103

11 Feng, J. *et al.* Imaging of Optically Active Defects with Nanometer Resolution. *Nano Lett.* **18**, 1739-1744 (2018). https://doi.org/10.1021/acs.nanolett.7b04819

12 Inglesby, M. K. & Zeronian, S. H. Diffusion coefficients for direct dyes in aqueous and polar aprotic solvents by the NMR pulsed-field gradient technique. *Dyes and Pigments* **50**, 3-11 (2001). https://doi.org/https://doi.org/10.1016/S0143-7208(01)00035-3

13 Comtet, J. *et al.* Direct observation of water-mediated single-proton transport between hBN surface defects. *Nature Nanotechnology* **15**, 598-604 (2020). https://doi.org/10.1038/s41565-020-0695-4

14 Ronceray, N. *et al.* Liquid-activated quantum emission from pristine hexagonal boron nitride for nanofluidic sensing. *Nature Materials* **22**, 1236-1242 (2023). https://doi.org/10.1038/s41563-023-01658-2

15 Shi, J. *et al.* Solution-phase sample-averaged single-particle spectroscopy of quantum emitters with femtosecond resolution. *Nature Materials* **23**, 1063-1069 (2024). https://doi.org/10.1038/s41563-024-01855-7

16 Smit, R., Tebyani, A., Hameury, J., van der Molen, S. J. & Orrit, M. Sharp zero-phonon lines of single organic molecules on a hexagonal boron-nitride surface. *Nature Communications* **14**, 7960 (2023). https://doi.org/10.1038/s41467-023-42865-4

17 Fumagalli, L. *et al.* Anomalously low dielectric constant of confined water. *Science* **360**, 1339-1342 (2018). https://doi.org/10.1126/science.aat4191

18 Sülzle, J. *et al.* Label-Free Imaging of DNA Interactions with 2D Materials. *ACS Photonics* **11**, 737-744 (2024). https://doi.org/10.1021/acsphotonics.3c01604

19 Gao, X. *et al.* Quantum Sensing of Paramagnetic Spins in Liquids with Spin Qubits in Hexagonal Boron Nitride. *ACS Photonics* **10**, 2894-2900 (2023). https://doi.org/10.1021/acsphotonics.3c00621

20 de Haas, T. *et al.* Charge Transfer-Induced Weakening of Vibronic Coupling for Single Terrylene Molecules Adsorbed onto Hexagonal Boron Nitride. *The Journal of Physical Chemistry Letters* **16**, 349-356 (2025). https://doi.org/10.1021/acs.jpclett.4c02899

21 Szalai, A. M. *et al.* Real-Time Structural Biology of DNA and DNA-Protein Complexes on an Optical Microscope. *bioRxiv*, 2023.2011.2021.567962 (2023). https://doi.org/10.1101/2023.11.21.567962

22 Lan, L. *et al.* Phase-Dependent Fluorescence Quenching Efficiency of MoS2 Nanosheets and Their Applications in Multiplex Target Biosensing. *ACS Applied Materials & Interfaces* **10**, 42009-42017 (2018). https://doi.org/10.1021/acsami.8b15677

23 Comtet, J. *et al.* Wide-Field Spectral Super-Resolution Mapping of Optically Active Defects in Hexagonal Boron Nitride. *Nano Lett.* **19**, 2516-2523 (2019). https://doi.org/10.1021/acs.nanolett.9b00178

24 Zhang, Z., Kenny, S. J., Hauser, M., Li, W. & Xu, K. Ultrahigh-throughput single-molecule spectroscopy and spectrally resolved super-resolution microscopy. *Nat. Methods* **12**, 935-938 (2015). https://doi.org/10.1038/nmeth.3528

25 Dong, B. *et al.* Super-resolution spectroscopic microscopy via photon localization. *Nature Communications* **7**, 12290 (2016). https://doi.org/10.1038/ncomms12290

26 Tran, T. T., Bray, K., Ford, M. J., Toth, M. & Aharonovich, I. Quantum emission from hexagonal boron nitride monolayers. *Nature Nanotechnology* **11**, 37-41 (2016). https://doi.org/10.1038/nnano.2015.242

27 Choi, S. *et al.* Engineering and Localization of Quantum Emitters in Large Hexagonal Boron Nitride Layers. *ACS Applied Materials & Interfaces* **8**, 29642-29648 (2016). https://doi.org/10.1021/acsami.6b09875

28 Haykal, A. *et al.* Decoherence of $V_B$ spin defects in monoisotopic hexagonal boron nitride. *Nature Communications* **13**, 4347 (2022). https://doi.org/10.1038/s41467-022-31743-0

29 Mendelson, N. *et al.* Identifying carbon as the source of visible single-photon emission from hexagonal boron nitride. *Nature Materials* **20**, 321-328 (2021). https://doi.org/10.1038/s41563-020-00850-y

30 Li, S. & Gali, A. Identification of an Oxygen Defect in Hexagonal Boron Nitride. *The Journal of Physical Chemistry Letters* **13**, 9544-9551 (2022). https://doi.org/10.1021/acs.jpclett.2c02687

31 Gao, X. *et al.* Single nuclear spin detection and control in a van der Waals material. *Nature* (2025). https://doi.org/10.1038/s41586-025-09258-7

32 Glushkov, E. *et al.* Engineering Optically Active Defects in Hexagonal Boron Nitride Using Focused Ion Beam and Water. *ACS Nano* **16**, 3695-3703 (2022). https://doi.org/10.1021/acsnano.1c07086

33 Xu, X. *et al.* Creating Quantum Emitters in Hexagonal Boron Nitride Deterministically on Chip-Compatible Substrates. *Nano Lett.* **21**, 8182-8189 (2021). https://doi.org/10.1021/acs.nanolett.1c02640

34 Zhang, O. *et al.* Six-dimensional single-molecule imaging with isotropic resolution using a multi-view reflector microscope. *Nature Photonics* **17**, 179-186 (2023). https://doi.org/10.1038/s41566-022-01116-6

35 Rimoli, C. V., Valades-Cruz, C. A., Curcio, V., Mavrakis, M. & Brasselet, S. 4polar-STORM polarized super-resolution imaging of actin filament organization in cells. *Nature Communications* **13**, 301 (2022). https://doi.org/10.1038/s41467-022-27966-w

36 Gould, T. J. *et al.* Nanoscale imaging of molecular positions and anisotropies. *Nat. Methods* **5**, 1027-1030 (2008). https://doi.org/10.1038/nmeth.1271

37 Backlund, M. P. *et al.* Simultaneous, accurate measurement of the 3D position and orientation of single molecules. *Proceedings of the National Academy of Sciences* **109**, 19087-19092 (2012). https://doi.org/doi:10.1073/pnas.1216687109

38 Curcio, V., Alemán-Castañeda, L. A., Brown, T. G., Brasselet, S. & Alonso, M. A. Birefringent Fourier filtering for single molecule coordinate and height super-resolution imaging with dithering and orientation. *Nature Communications* **11**, 5307 (2020). https://doi.org/10.1038/s41467-020-19064-6

39 Hulleman, C. N. *et al.* Simultaneous orientation and 3D localization microscopy with a Vortex point spread function. *Nature Communications* **12**, 5934 (2021). https://doi.org/10.1038/s41467-021-26228-5

40 Thorsen, R. Ø., Hulleman, C. N., Rieger, B. & Stallinga, S. Photon efficient orientation estimation using polarization modulation in single-molecule localization microscopy. *Biomed. Opt. Express* **13**, 2835-2858 (2022). https://doi.org/10.1364/BOE.452159

41 Backer, A. S., Lee, M. Y. & Moerner, W. E. Enhanced DNA imaging using super-resolution microscopy and simultaneous single-molecule orientation measurements. *Optica* **3**, 659-666 (2016). https://doi.org/10.1364/OPTICA.3.000659

42 Hofmann, A., Schmid, M. & Brütting, W. The Many Facets of Molecular Orientation in Organic Optoelectronics. *Advanced Optical Materials* **9**, 2101004 (2021). https://doi.org/https://doi.org/10.1002/adom.202101004

43 Karedla, N. *et al.* Simultaneous Measurement of the Three-Dimensional Orientation of Excitation and Emission Dipoles. *Physical Review Letters* **115**, 173002 (2015). https://doi.org/10.1103/PhysRevLett.115.173002

44 Tenopala-Carmona, F. *et al.* Orientation distributions of vacuum-deposited organic emitters revealed by single-molecule microscopy. *Nature Communications* **14**, 6126 (2023). https://doi.org/10.1038/s41467-023-41841-2

45 Mayner, E. *et al.* Monitoring Electrochemical Dynamics through Single-Molecule Imaging of hBN Surface Emitters in Organic Solvents. *ACS Nano* **18**, 27401-27410 (2024). https://doi.org/10.1021/acsnano.4c07189

46 Zhang, O. & Lew, M. D. Quantum limits for precisely estimating the orientation and wobble of dipole emitters. *Physical Review Research* **2**, 033114 (2020). https://doi.org/10.1103/PhysRevResearch.2.033114

47 Ding, T., Wu, T., Mazidi, H., Zhang, O. & Lew, M. D. Single-molecule orientation localization microscopy for resolving structural heterogeneities between amyloid fibrils. *Optica* **7**, 602-607 (2020). https://doi.org/10.1364/OPTICA.388157

48 Agrawal, A., Quirin, S., Grover, G. & Piestun, R. Limits of 3D dipole localization and orientation estimation for single-molecule imaging: towards Green's tensor engineering. *Optics Express* **20**, 26667-26680 (2012). https://doi.org/10.1364/OE.20.026667

49 Backer, A. S. & Moerner, W. E. Determining the rotational mobility of a single molecule from a single image: a numerical study. *Optics Express* **23**, 4255-4276 (2015). https://doi.org/10.1364/OE.23.004255

50 Backer, A. S. & Moerner, W. E. Extending Single-Molecule Microscopy Using Optical Fourier Processing. *The Journal of Physical Chemistry B* **118**, 8313-8329 (2014). https://doi.org/10.1021/jp501778z

51 Descloux, A., Grußmayer, K. S. & Radenovic, A. Parameter-free image resolution estimation based on decorrelation analysis. *Nat. Methods* **16**, 918-924 (2019). https://doi.org/10.1038/s41592-019-0515-7

52 Yang, Y. *et al.* Electrochemically controlled blinking of fluorophores for quantitative STORM imaging. *Nature Photonics* (2024). https://doi.org/10.1038/s41566-024-01431-0
53 Yu, M., Yim, D., Seo, H. & Lee, J. Electrical charge control of h-BN single photon sources. *2D Materials* **9**, 035020 (2022). https://doi.org/10.1088/2053-1583/ac75f4
54 Wong, D. *et al.* Characterization and manipulation of individual defects in insulating hexagonal boron nitride using scanning tunnelling microscopy. *Nature Nanotechnology* **10**, 949-953 (2015). https://doi.org/10.1038/nnano.2015.188
55 Shin, D. H. *et al.* Diffusion of DNA on Atomically Flat 2D Material Surfaces. *ACS Nano* **19**, 21307-21318 (2025). https://doi.org/10.1021/acsnano.4c16277
56 Reinhardt, S. C. M. *et al.* Ångström-resolution fluorescence microscopy. *Nature* **617**, 711-716 (2023). https://doi.org/10.1038/s41586-023-05925-9

## Acknowledgments

The authors thank Dr. Yunfei Teng for his help with the figure rendering of the experimental concept.

## Funding Statement

W.G., T.C., N.R., E.M. and A.R. acknowledge funding from the European Research Council Grant (No. 101020445—2D-LIQUID, and No. 101248059—MicroSwiss). The authors are supported by the EPFL Center for Imaging through its 2022 Call for Interdisciplinary Projects in Imaging. K.W. and T.T. acknowledge support from the JSPS KAKENHI (Grant Numbers 21H05233 and 23H02052) and World Premier International Research Center Initiative (WPI), MEXT, Japan.

## Author contributions Statement

A.R., W.G., and N.R. designed the experiments. W.G. built the microscope and programmed the code. W.G. performed experiments with suggestions from A.R., N.R., T.C., and E.M.. W.G. and T.C. analyzed the data. W.G. wrote the manuscript with the help of A.R. and input from all authors. K.W. and T.T. contributed the materials. A.R. supervised the project. All authors discussed the results and commented on the manuscript.

## Competing Interests Statement

The authors declare no conflicts of interest.

# Figure Captions

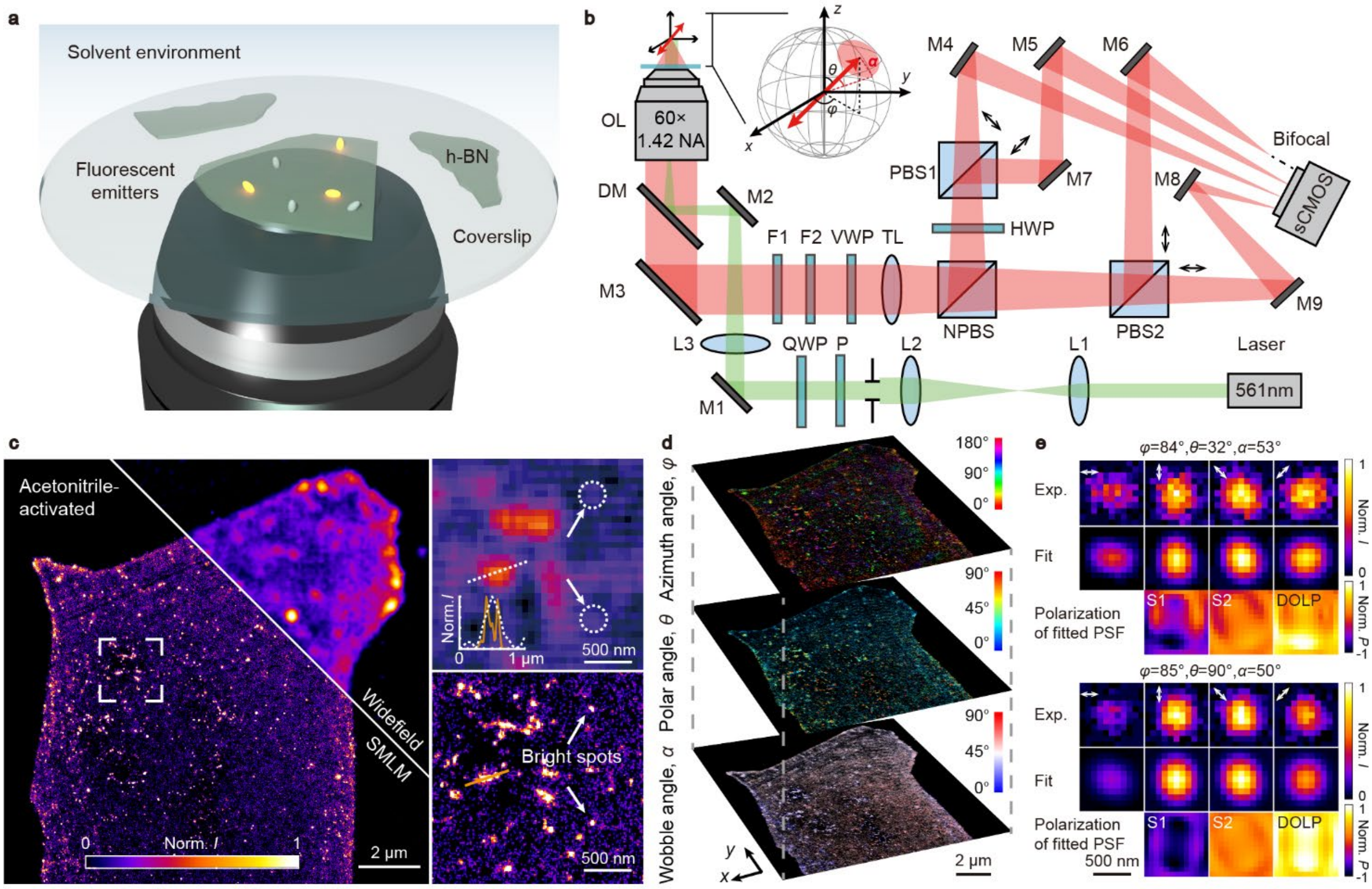

**Fig. 1 Molecular dynamics of emitters at h-BN surface.** (a) The illustration of the generation of fluorescent emitters at the h-BN and solvent interface. (b) The schematic diagram of polarized single-molecule localization microscopy (pSMLM) setup. L1-L3: lens; P: polarizer; QWP: quarter waveplate; M1-M9: mirrors; OL: objective lens; DM: dichroic mirror; F1-F2: filters; VWP: variable waveplate; TL: tube lens; NPBS: non-polarizing beamsplitter; PBS1-PBS2: polarizing beamsplitters; HWP: half waveplate. (c) The comparison of SMLM and widefield images. The SMLM records more interaction activity and displays them as bright spots. Number of independent replicates for different samples $n$>3; (d) Super-resolution rendering of the 3D angular parameters of the emitters, which demonstrates their dynamics and preferences on the h-BN surface. (e) Two representative PSFs with large wobble angles. From top to bottom are the experimental results, fitted results, and polarization state (Stokes vectors and degree of linear polarization). The asymmetry of the Stokes image reflects the change of polar angle. Source data are provided as a Source Data file

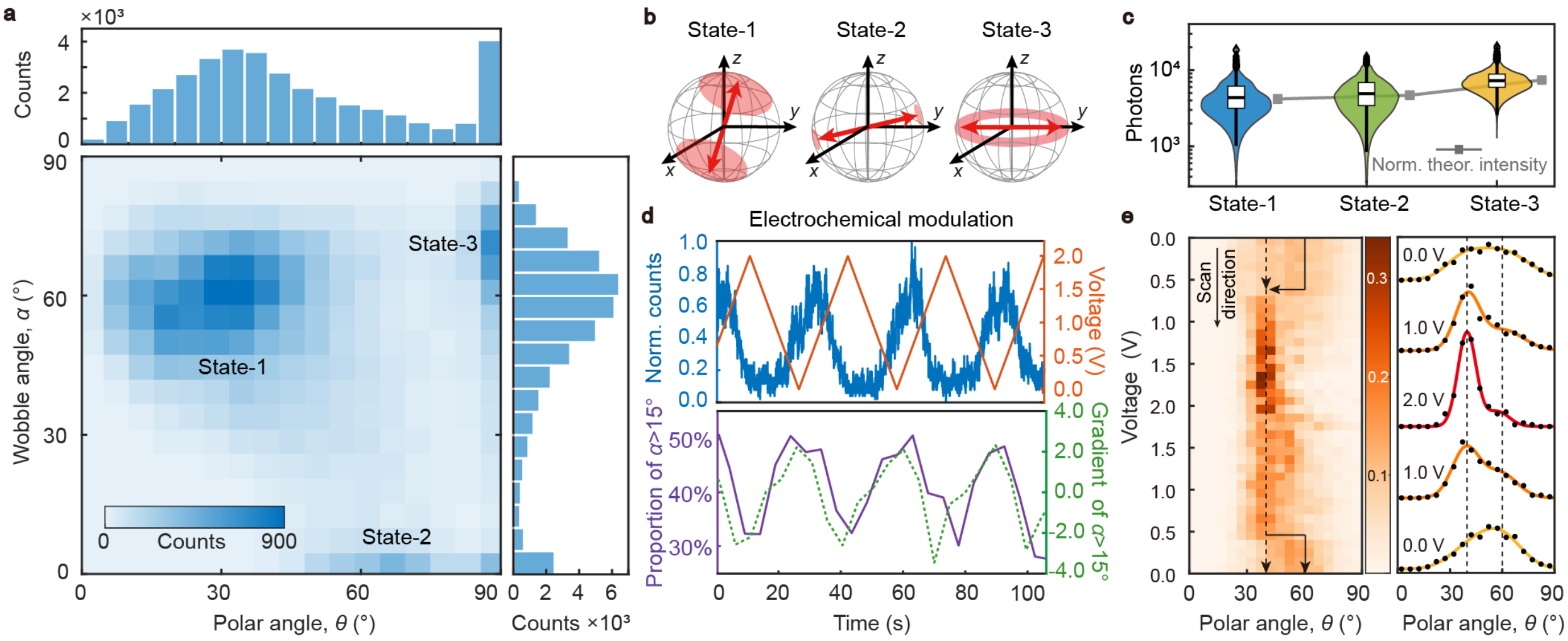

**Fig. 2 Breaking of axial symmetry demonstrates single-molecule interaction with the h-BN surface.** (a) The density map of polar angle versus wobble angle shows three clusters; (b) The arrows indicate the orientations of the three emitter states in spherical coordinates, and the shaded areas indicate the wobbling range; (c) The light-intensity histograms of three states, which are in ascending order with the polar angle and agree with the normalized theoretical values. The localization points are 79001,6783, and 4344 respectively. The simulation data is derived from the center values of the three states. For the box plots, the center line represents the median, and the bounds of the box indicate the 25th and 75th percentiles. The whiskers extend to the minimum and maximum values within 1.5 times the interquartile range (IQR), and outliers are plotted individually as points. (d) Regulating effect of the vertical potential on the emitters. In the top panel, the orange curve is a triangle wave that is applied to the ITO electrode below the sample (against the Pt electrode) and the blue curve is the normalized localization counts. The bottom panel shows the proportion (violet) and gradient (dash green) of emitters with $\alpha$>15° (i.e., state-1 and -3) to the total number of localizations under current voltages; (e) Shift on the polar angle of state-2 under the electrochemical modulation. The pixels in each row represent the state-2 histogram at the current voltage. The polar angle distributions for five typical conditions are shown on the right. Source data are provided as a Source Data file

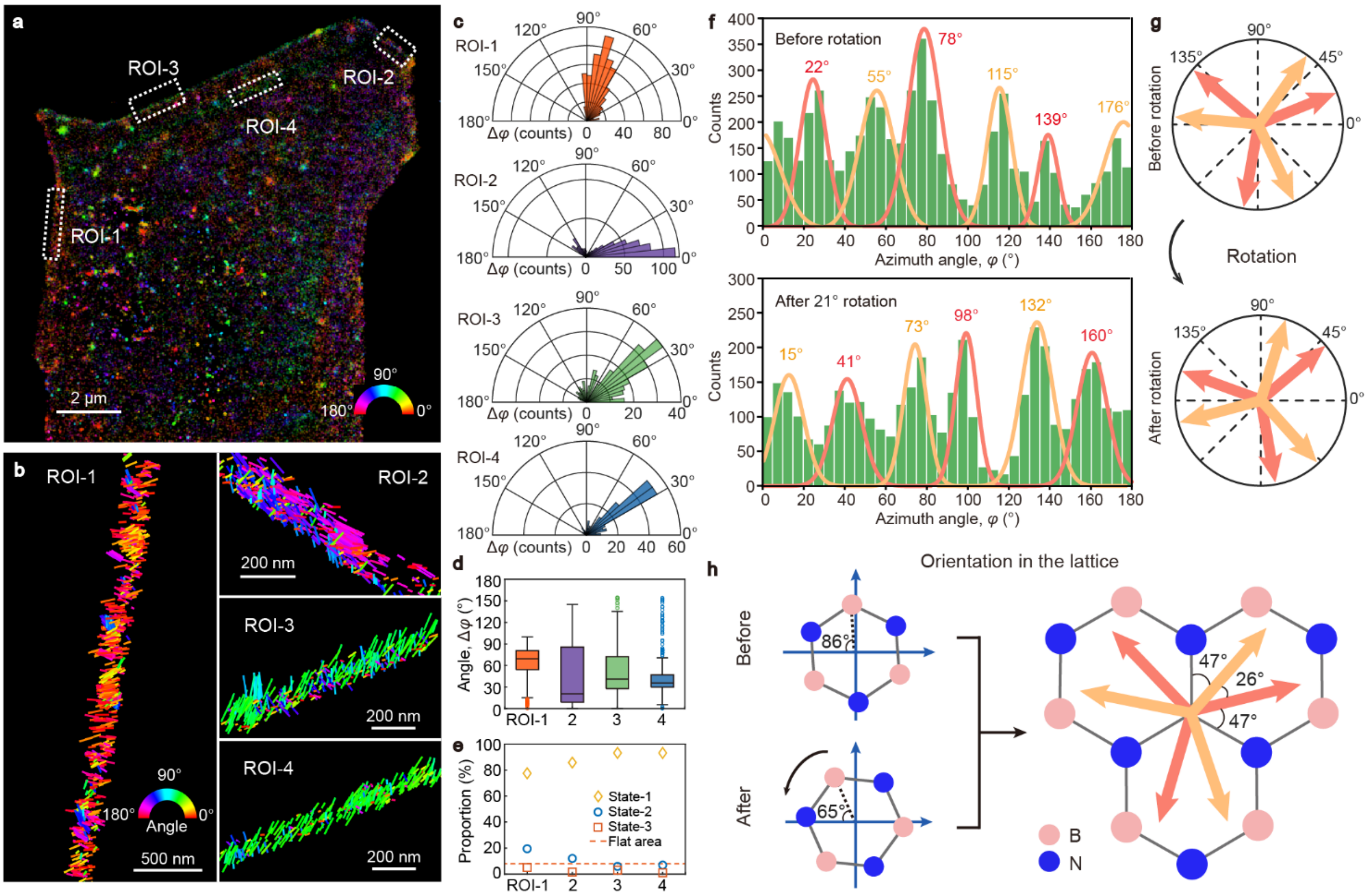

**Fig. 3 Azimuth angle of emitters on the h-BN surface.** (a) The SMLM image is rendered according to azimuth angle of emitters and temporally stacked. Number of independent replicates for different samples *n*>3; (b) Zoomed-in view of the four regions of interest in (a) with the azimuthal orientation marked with sticks and colors; (c) the angular distribution of $\Delta\varphi=|\varphi_{edge}-\varphi_{emitters}|$ in four ROIs; (d) Boxplot of the angle differences of four ROIs. The localization points are 582, 597, 306, and 251 respectively. For the box plots, the center line represents the median, and the bounds of the box indicate the 25th and 75th percentiles. The whiskers extend to the minimum and maximum values within 1.5 times the interquartile range (IQR), and outliers are plotted individually as points. (e) The fraction of the three states at the edge of the crystal (P1+P2+P3=1 in each ROI) and the dashed line is the share of state-3 in the flat region. (f) Azimuth histogram of the state-2 emitters in the flat crystal area. Top: Before rotation, Bottom: After 21° horizontal rotation. (g) Azimuthal angles are restored to full space based on crystal symmetry. (h) Superposition of the azimuthal angle of the emitters with the crystal lattice. Source data are provided as a Source Data file

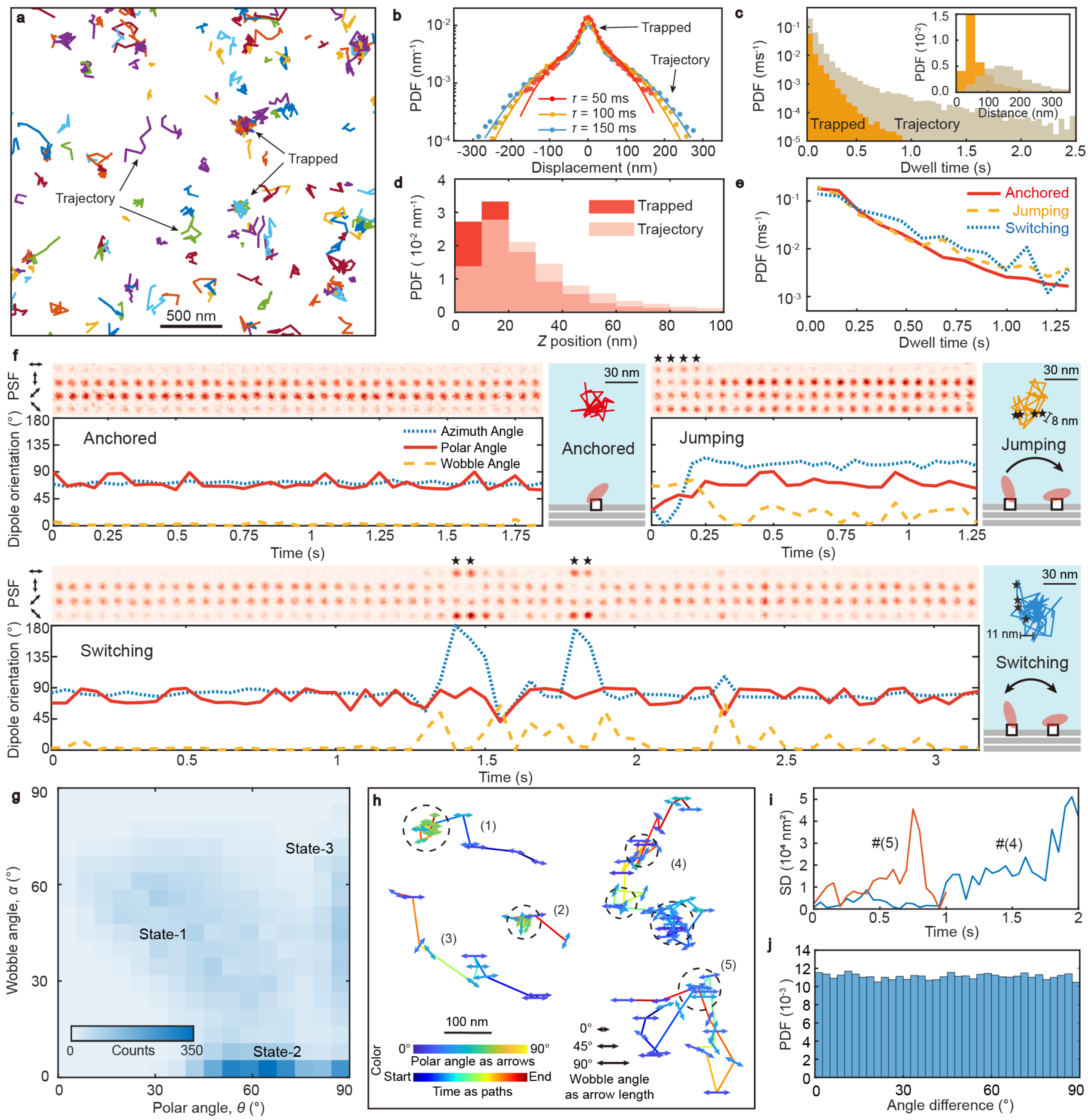

**Fig. 4 Molecule dynamics at solid-liquid interface.** (a) Trajectory map of events that are longer than 4 frames (200 ms). (b) The 1D displacement of all events with different lag time: 50, 100, and 150 ms. (c) The dwell time for trapped and trajectory events. The insert figure shows the distribution of traveling distance for two types of events. (d) The *Z*-position of trapped emitters and hopping emitters. (e) The dwell time for the three subcategories of trapped events. (f) Three subcategories of trapped events: Anchored, jumping and switching. In the left panel of each event are the experimental PSFs and the fitted dipole orientation. In the right panel is the 2D trajectory and schematic illustration. (g) The 2D density plot shows that the emitter in state-2 dominates the trapped events. (h) Three basic trajectory events and their combination. The orientation of the arrow indicates the azimuth angle of the emitters, the color indicates the polar angle and the length indicates the wobble angle. (i) the square displacement (SD) of #4 and #5 in (h). (j) The histogram of the angle difference between the trajectory direction and the azimuth angle of each emitter. Source data are provided as a Source Data file